# Associating Frailty and Dynamic Dysregulation between Motor and Cardiac Autonomic Systems

Patricio Arrué[1], Kaveh Laksari[1,2], Nima Toosizadeh*[1,3,4]

[1]Department of Biomedical Engineering, University of Arizona, Tucson, AZ, United States.
[2]Department of Aerospace and Mechanical Engineering, University of Arizona, Tucson, AZ, United States.
[3]Arizona Center on Aging (ACOA), Department of Medicine, University of Arizona, Tucson, AZ, United States.
[4]Division of Geriatrics, General Internal Medicine and Palliative Medicine, Department of Medicine, University of Arizona, Tucson, AZ, United States.

## Abstract

Frailty is a geriatric syndrome associated with the lack of physiological reserve and consequent adverse outcomes (therapy complications and death) in older adults. Recent research has shown associations between heart rate (HR) dynamics (HR changes during physical activity) with frailty. The goal of the present study was to determine the effect of frailty on the interconnection between motor and cardiac systems during a localized upper-extremity function (UEF) test. Fifty-six older adults aged 65 or older were recruited and performed the UEF task of rapid elbow flexion for 20-seconds with the right arm. Frailty was assessed using the Fried phenotype. Wearable gyroscopes and electrocardiography were used to measure motor function and HR dynamics. Using convergent cross-mapping (CCM) the interconnection between motor (angular displacement) and cardiac (HR) performance was assessed. A significantly weaker interconnection was observed among pre-frail and frail participants compared to non-frail individuals ($p$<0.01, effect size=0.81±0.08). Using logistic models pre-frailty and frailty were identified with sensitivity and specificity of 82% to 89%, using motor, HR dynamics, and interconnection parameters. Findings suggested a strong association between cardiac-motor interconnection and frailty. Adding CCM parameters in a multimodal model may provide a promising measure of frailty.

## Introduction

Frailty is an aging syndrome related to low physiological reserves in organs and systems and is associated with increased risk of hospitalization, adverse therapy outcomes, disability, and mortality (1). Muscle loss (sarcopenia), and weakness (dynapenia) are the main symptoms of frailty (2), which are triggered by metabolic and hormonal derangements (3–6), and the so called "heightened inflammatory state" (7), caused by excessive levels of C-reactive protein (CRP), proinflammatory cytokines interleukin 6 (IL-6), and white blood cells and tumor necrosis factor-alpha (TNFalpha) (7–9). Consequently, frailty is highly associated with a decrease in motor function performance (10). Furthermore, frailty is associated with an impaired cardiac autonomic nervous system (ANS) because of alterations in the action potential on the sinoatrial node myocytes, which impacts the cardiac function and the heart rate variability (HRV) (11). While research showed alterations in several physiological systems, the association between frailty and dynamic interconnection between cardiac and motor systems is still unclear. Indeed, the human body is a complex network of several physiological systems, where intricate dynamics exist between these systems to maintain homeostasis (12,13). Accurate identification of the level of physiological reserve requires a collection of information across multiple physiological systems (12–15), rather than only system-specific evaluations. To explore the extent of dynamic behaviors within and across physiological systems, principles of network physiology has been introduced. The concept of network physiology claims that dysregulation of interactions between physiological systems leads to loss of resilience and the ability to recover from stressors (14), which is inherent to the concept of frailty.

We have previously developed a methodology for assessing frailty that incorporates an upper-extremity function (UEF) and corresponding heart rate (HR) response to physical activity. The UEF test consists of repetitive and rapid elbow flexion and extension (16), during which several kinematics features representing dynapenia are measured using motion sensors (1). Since UEF involves upper-extremity motion, it is feasible to perform for bedbound patients and where walking tests are difficult for frail older adults. In our recent research we showed that HR dynamics, measured by changes in HR due to the UEF physical function (i.e., HR dynamics), were significantly associated with frailty (17). Combining UEF motor and cardiac functions, we were able to identify frailty with higher accuracy compared to models including each of the motor or HR dynamics parameters separately (18). Nevertheless, it is unclear whether frailty can influence the interconnection between motor and HR dynamics, and whether applying interconnection measures improve frailty identification.

The goal of the current study was to determine the effect of frailty on the interconnection between motor and cardiac systems. Build upon our previous research, the main hypothesis was that due to frailty, a weaker interconnection would exist between motor and HR performance. Recently, the concept of interconnection assessment within different physiological systems has gained attention (19–24). Granger causality is a classical approach that identifies causality between variables based on the removal of one to determine the predictability of the other variable (25), but its usage is limited to linear systems that have stationary behaviors, or for those in which variables are strongly coupled (26). In contrast, convergent cross-mapping (CCM) assesses the non-linear directional interactions of variables in a complex dynamic system, based on state-space reconstruction of time series collected from each system (27). The secondary hypothesis was that the accuracy of frailty identification would be improved using additional CCM interconnection parameters compared to models incorporating each of motor and HR parameters individually.

## Methods

*Participants*

Older adult participants (≥65 years) were recruited between October 2016 and March 2018. Participants were recruited from primary, secondary, and tertiary health care settings such as primary and community care providers, assisted living facilities, retirement homes, and aging service organizations. The inclusion criteria were 1) being 65 years or older; 2) the ability to walk a minimum distance of 4.57 m (15 ft) for frailty assessment; and 3) the ability to read and sign an informed consent. The exclusion criteria were: 1) severe motor disorders (Parkinson's disease, multiple sclerosis, or recent stroke); 2) severe upper-extremity disorders (e.g., elbow bilateral fractures or rheumatoid arthritis); 3) cognitive impairment identified by a Mini Mental State Examination (MMSE) score ≤ 23 (28); 4) terminal illness; 5) diseases/treatments that can bias the HR measurements (including arrhythmia and use of pacemaker); and 6) usage of β-blockers or similar medications that can influence HR. Written informed consent was obtained from all participants. The study was approved by the University of Arizona Institutional Review Board. All research was performed in accordance with the relevant guidelines and regulations, according to the principles expressed in the Declaration of Helsinki (29).

*Frailty assessment and clinical measures*

The frailty assessment was executed using the five-component Fried phenotype as the gold standard (1). The Fried phenotype considers five criteria: 1) unintentional weight loss of 4.54 kg (10 pounds) or more in the previous year; 2) grip strength weakness (adjusted with body mass index (BMI) and sex); 3) slowness based on the required time to walk 4.57 m (15 ft) (adjusted with height and sex); 4) self-reported exhaustion based on a short two-question version of Center for Epidemiological Studies Depression (CES-D); and 5) self-reported low energy expenditure based on a short version of Minnesota Leisure Time Activity Questionnaire (30). Participants were categorized into three frailty groups, which were non-frail if they met none of the criteria, pre-frail if they met one or two criteria, and frail if they met three or more criteria. Clinical measures collected included: 1) MMSE and Montreal Cognitive Assessment (MoCA) for cognition (28,31); 2) comorbidity based on Charlson Comorbidity Score (CCI) (32); and 3) depression using Patient Health Questionnaire (PHQ-9) (33). These measures were considered as adjusting variables in the statistical analysis because they could potentially influence physical activity and the cardiovascular system performance.

*UEF test*

After frailty assessment and clinical measures, participants were asked to sit on a chair and rest for two minutes to regain normal resting status. Participants then performed the UEF task of elbow flexion-extension as quickly as possible for 20 seconds with the right arm. After the UEF task, participants rested on the chair for another two minutes. We have shown that UEF results are similar on both left and right hands (34). Before the test, participants practiced the UEF test with their non-dominant arm to become familiar with the protocol. The protocol was explained to participants and using exact same verbal instruction they were encouraged only once, before elbow flexion, to do the task as fast as possible. Wearable motion sensors (triaxial gyroscope sensors, BioSensics LLC, Cambridge, MA, sampling frequency=100 Hz; Figure 1A) were used to measure forearm and upper arm motion, and ultimately the elbow angular velocity. Angular velocity data from gyroscopes were filtered using first-order high pass butter-worth filter with a cutoff of 2.5 Hz. Maximums and minimums of the angular velocity signal were detected, and subsequently, elbow flexion cycles were identified. Motor performance was assessed to represent: 1) slowness based on speed of elbow flexion; 2) flexibility based on range of motion, 3) weakness based on strength of upper-extremity muscles; 4) speed variability and motor accuracy; 5) fatigue based on reduction in speed during the 20-second task, and 6) number of flexion cycles. A sub-score was assigned for each of those features, determined previously based on multivariable ordinal logistic models, with the Fried frailty categories as the dependent variable and UEF parameters plus demographic information as independent variables (17). The normalized UEF motor score from zero (resiliency) to one (extreme frailty) was computed as the sum of sub-scores corresponding to performance results and demographic information (i.e., BMI) (16). More details about UEF validation, repeatability, and the normalized score are explained in previous research (16,34,35).

HR was recorded using a wearable ECG device with two electrodes and one built-in accelerometer (360° eMotion Faros, Mega Electronics, Kuopio, Finland; ECG sampling frequency=1000 Hz and accelerometer sampling frequency=100 Hz; Figure 1A). One ECG electrode was placed on the upper mid-thorax and the other one inferior to the left rib cage. The placement of the electrodes on the left chest would minimize the movement artifacts due to UEF test with the right arm. ECG data was analyzed for 20 seconds of baseline, 20-second UEF, and 30 seconds of recovery. RR intervals (successive R peaks of the QRS signal) were computed using the Pan-Tompkins algorithm (36). The automated peak detection process was manually inspected by two researchers (PA and NT). Previously two types of HR parameters were extracted, one representing baseline HR and HR variability (HRV), and one representing HR dynamics (changes in HR during UEF and HR recovery after the task) (37). Briefly, HR dynamics included time to reach maximum and minimum HR, as well as percent increase and decrease in HR during activity and recovery periods, respectively. In addition to previously developed parameters, in the current study, the interconnection between motor and HR data were assessed using CCM.

*CCM analysis*

We quantitatively assessed the directional nonlinear interactions between HR and motor data using CCM. An overview of the method is summarized in Figure 1. CCM tests whether a historical trace of HR can predict motor performance (or inversely, whether a historical trace motor performance can predict HR). To calculate the CCM, we first created evenly sampled data of synchronized HR and motor function with a sampling frequency of 10Hz, using spline interpolation (Figure 1B). Each HR data point represents average HR values over 0.1 seconds. Corresponding motor data represent the angular displacement travelled during each 0.1 second of UEF. For calculating motor performance, motor function $M_f$ was defined by:

$$M_{f_i} = \int_{t_i}^{t_i+0.1} \omega_e dt, \tag{1}$$

where $\omega_e$ represents the angular velocity of the elbow.

Taken's embedding theorem generally guarantees that the space state of a dynamic system could be represented from a single-observed time series $X$ as an $E$-dimensional manifold (38). The shadow or reconstructed manifold, denoted by $M_X$, consists of an $E$-dimensional data with lagged coordinates ($\tau$) of the variable:

$$M_X = \langle X(t), X(t-\tau), X(t-2\tau) \ldots X(t-(E-1)\tau) \rangle. \tag{2}$$

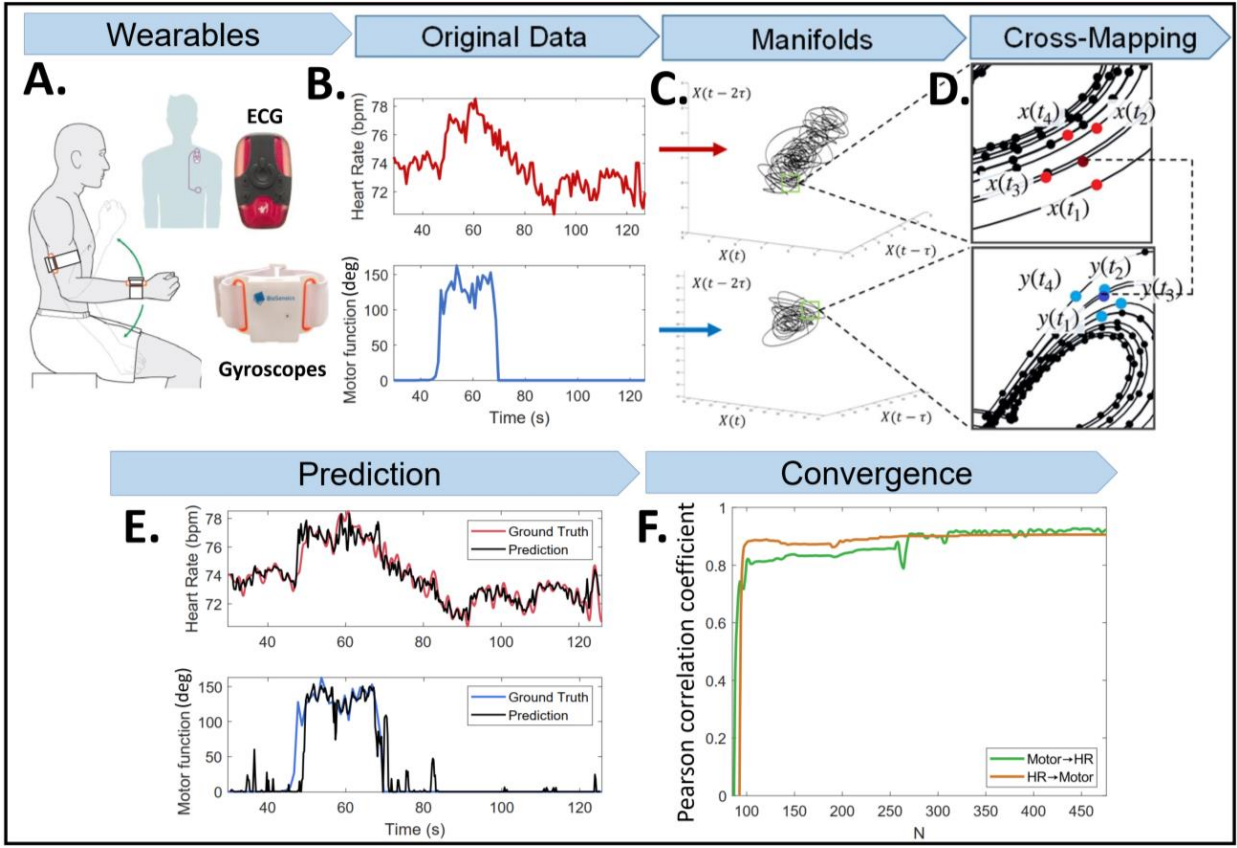

**Figure 1:** Overview of the CCM method to assess interconnection between motor and HR data: A) Wearable devices (gyroscopes) to obtain angular velocity and ECG during the UEF physical task; B) Motor performance and HR extraction; C) CCM shadow attractor manifolds on time-lagged coordinate systems; D) Prediction of HR from motor function and vice-versa in a time point (dark red and dark blue dots, respectively) using a distance-based weighted average of neighbors (bright red and bright blue dots); E) Comparison between predicted motor (or HR) data and ground truth; and F) Convergence curves of Pearson correlation coefficient between predicted and ground truth as a function of library length (data points used for developing manifolds).

Subsequently, we reconstructed $E$-dimensional manifolds from each of these two time series (38) (Figure 1C). Dimension ($E$) of four was selected based on the average false nearest neighbor approach (39). A time lag ($\tau$) of 1 second was used for analysis based on the delayed mutual information method (40). We predicted one time series (e.g., motor function) by historical records of the other signal (e.g., shadow manifold of HR) using a k-nearest neighbor technique. For a dimension $E$, we determined $E + 1$ nearest neighbors and identified indices of each data points in manifolds ($M_X$). Using these indices for one manifold (e.g., motor data $X(t)$), we found corresponding neighbors in the second manifold (e.g., HR data $Y(t)$) (Figure 1D), and then predicted $X(t)$ to $\hat{Y}(t)$ as the weighted mean of $E + 1$ points in the second manifold (41):

$$\hat{Y}(t) = \sum_{i=1}^{E+1} w_i Y(t_i), \tag{3}$$

where $w_i$ weights are calculated based on the Euclidean distances between $M_Y$ and its $i^{th}$ nearest neighbor on $X(t)$.

The Pearson correlation coefficient and the normalized root-mean-square-error (NRMSE) between the predicted and original time series were calculated to assess the strength of interconnections (Figure 1E). NRMSE was calculated by normalizing the RMSE between the predicted and the ground truth with respect to the standard deviation of observations. As documented in previous studies, the correlation coefficient is expected to increase with increasing the time-series length (i.e., library length, Figure 1F). For the current study, the correlation and NRMSE values were calculated at the maximum library length (Figure 1F).

*Statistical analysis*

Analysis of variance (ANOVA) models were used to evaluate the differences in demographic information between the frailty groups, except for sex. Instead, the chi-square ($\chi^2$) test was used to assess the difference in sex categories among frailty groups. CCM parameters were compared between frailty groups using multivariable ANOVA models; age, sex, and BMI were considered as adjusting variables since they have been previously associated with motor and cardiac performance and frailty (16,42–44). Cohen's effect size (*d*) was estimated. ANOVA analyses for comparing CCM parameters across frailty groups were repeated with clinical measures with significant association with frailty as covariates. To assess the additional value of interconnection measures compared to previous models with individual motor and HR parameters, logistic regression models were implemented with frailty as the dependent variable and HR, motor, and CCM parameters as independent parameters. A stepwise parameter selection based on Akaike information criterion (AIC) values was implemented to identify independent predictive variables. The area under the curve (AUC) with 95% confidence interval (CI) was calculated using receiver operator characteristics (ROC) curves for each predicting model. Statistical analyses were done using JMP (Version 16, SAS Institute Inc., Cary, NC, USA), and statistical significance was concluded when *p*<0.05.

## Results

*Participants and clinical measures*

Fifty-six participants were recruited for the study, including 12 non-frail (age=76.92±7.32 years), 40 pre-frail (age=80.53±8.12 years), and four frail individuals (age=88.25±4.43 years). Of note, due to the small number of frail participants (n=4), frail and pre-frail groups were merged for the statistical analysis. A summary of demographics is presented in Table 1. There was no significant difference in demographic parameter between the frailty groups (*p*>0.10). Among clinical measures, CCI comorbidity and PHQ-9 depression scores were significantly different between frailty groups (*p*<0.03, Table 1).

*CCM analysis*

Significant effects of frailty on CCM correlation values were observed for interconnections in both directions, i.e., predicting HR time series based on motor function (motor to HR) and predicting motor function based on HR (HR to motor) as reported in Table 2, Figure 2 and 3. Pre-frail/frail older adults showed smaller correlations in CCM for both directions, compared to non-frail older adults. There was also a significant effect of frailty on NRMSE values (*p*<0.05); for both motor and HR CCM predictions, NRMSE values were significantly smaller among non-frail compared to pre-frail/frail (*p*<0.05).

**Table 1.** Demographic information and clinical measures of participants.

| Variables | Non-frail (n=12) | Pre-frail/Frail (n=44) | *p*-value (effect size) |
|---|---|---|---|
| Female, n (% of the group) | 7 (42%) | 34 (77%) | 0.20 |
| Age, year (SD) | 76.92 (7.32) | 81.23 (8.14) | 0.10 (0.54) |
| Height, cm (SD) | 164.36 (9.13) | 164.23 (10.27) | 0.97 (0.01) |
| Weight, kg (SD) | 66.58 (14.69) | 75.53 (19.56) | 0.15 (0.48) |
| Body mass index, kg/m$^2$ (SD) | 24.67 (5.55) | 27.74 (5.71) | 0.10 (0.54) |
| MMSE score, 0-30 (SD) | 29.67 (0.65) | 29.14 (1.34) | 0.19 (0.53) |
| MoCA score, 0-30 (SD) | 26.25 (3.08) | 24.88 (2.80) | 0.15 (0.48) |
| CCI score, 0-29 (SD) | 1.42 (1.78) | 3.86 (2.89) | <0.01* (0.91) |
| PHQ-9 score, 0-30 (SD) | 0.42 (0.51) | 2.35 (2.89) | 0.03* (0.75) |
| **Fried criteria, n (% of the group)** | | | |
| Weight loss | 0 | 1 (2%) | |
| Weakness | 0 | 18 (41%) | |
| Slowness | 0 | 34 (77%) | |
| Exhaustion | 0 | 7 (16%) | |
| Low energy | 0 | 8 (18%) | |

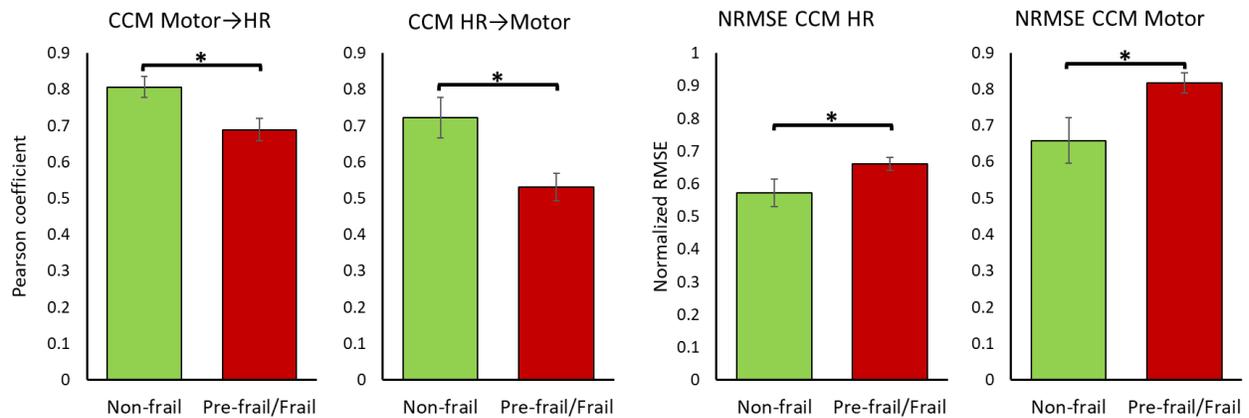

**Figure 2.** CCM parameters and NRMSE across frailty groups. A significant between group difference is identified by the asterisk (*p*<0.05).

Within the stepwise regression analysis, UEF score, HR percent increase, and CCM Motor-to-HR parameters were selected as independent predictors of frailty categories (non-frail vs. pre-frail/frail). Using these three parameters, pre-frailty/frailty was predicted with an AUC, sensitivity, and specificity of 0.91, 0.89, and 0.83 (Table 3, Figure 4), which had a 7% higher AUC than models that included only individual motor or HR parameters as predictors.

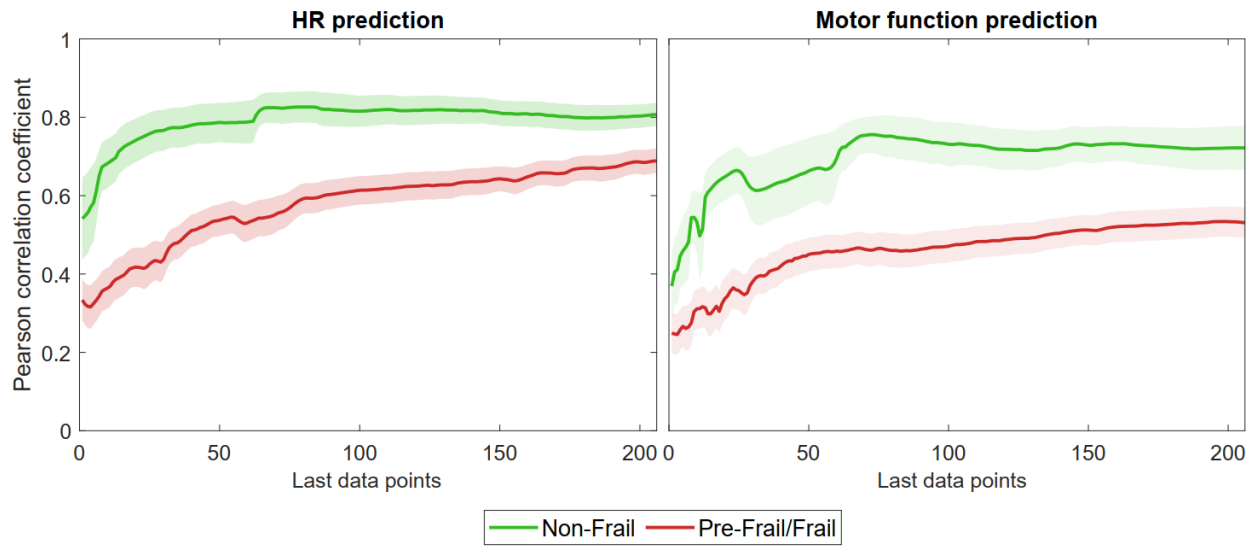

**Figure 3.** Convergence curves distribution for CCM predictions. Solid lines represent the average across each group at each library length and shaded regions show the standard error.

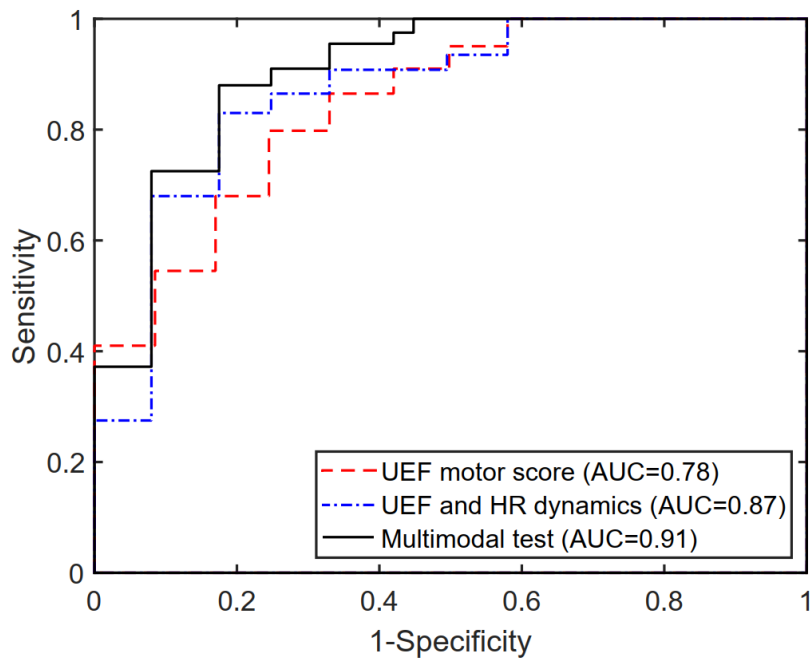

**Figure 4.** The area under the receiver operator characteristics (ROC) curve for the UEF score, the previous multimodal test (HR + UEF score), and the current model incorporating the CCM parameter.

**Table 2.** Differences in UEF, HRV, and CCM features across frailty groups. A significant association is represented by the asterisk.

| Parameters | Non-frail (n=12) | Pre-frail/Frail (n=44) | *P*-value (Effect size) |
|---|---|---|---|
| **UEF motor score** | | | |
| UEF motor score, 0-1 (SD) | 0.32 (0.18) | 0.53 (0.23) | 0.01* (0.90) |
| **HR dynamics parameters** | | | |
| HR mean, bpm (SD) | 71.52 (11.38) | 77.59 (15.97) | 0.23 (0.43) |
| Time to peak HR, seconds (SD) | 16.84 (6.46) | 16.12 (5.62) | 0.43 (0.28) |
| HR recovery time, seconds (SD) | 13.71 (6.22) | 14.04 (5.76) | 1.00 (0.001) |
| HR percent increase, % (SD) | 19.28 (7.55) | 10.29 (4.79) | <0.01*(1.49) |
| HR percent decrease, % (SD) | 15.24 (7.65) | 8.13 (3.97) | <0.01*(1.25) |
| **CCM parameters** | | | |
| Correlation Motor→HR (SD) | 0.81 (0.10) | 0.69 (0.21) | 0.03*(0.77) |
| Correlation HR→Motor (SD) | 0.72 (0.19) | 0.53 (0.26) | 0.01*(0.89) |
| NRMSE HR, % (SD) | 57.28 (14.47) | 66.10 (13.26) | 0.02*(0.86) |
| NRMSE Motor, % (SD) | 65.82 (21.67) | 81.68 (19.00) | 0.04*(0.73) |

*UEF: upper extremity function; SD: standard deviation; HR: heart rate; bpm: beats per minute; CCM: convergent cross-mapping; NRMSE: normalized root mean square error.*

## Discussion

*Effect of frailty on system interconnections*

As hypothesized, a significantly weaker interconnection between motor and cardiac systems was observed among pre-frail and frail older adults compared to non-frail individuals (Table 2 and Figure 2). Indeed, these results are consistent with expected changes due to aging-related physiological dysregulation. Autonomic nervous system (ANS) regulates heart activity during exercise by signals from the central nervous system (45) and feedback mechanisms from the exercise pressor reflex (group III and IV muscle afferents) (46) and the arterial baroreflex, which controls blood pressure and consequently cardiac output (47). Previous studies have shown that exercise pressor reflex is impacted by aging (48–51), which would potentially alter the interconnection between the motor and cardiac systems. Nevertheless, the effect is still controversial and further research is needed to fully understand this interconnection pathway. One potential explanation is that frailty leads to an altered control of motor to cardiac system by affecting exercise pressor reflex. Nevertheless, this hypothesis should be investigated in future research.

In addition to exercise pressor reflex, the observed weaker CCM values among pre-frail and frail older adults may be explained by the general concepts of homeostatic physiological dysregulation and heightened inflammatory state (7,52). In this regard, aging and more specifically frailty can be caused by breakdowns of key regulatory processes and excessive increase of immune factors, leading to the loss of homeostasis and functional impairment (3,5–7,9). Different methods have been used previously to identify physiological dysregulation, such as Mahalanobis multivariate statistical distance and principal component analysis. Mahalanobis multivariate statistical distance is a multivariate model built to assess dysregulation within relevant blood-based biomarkers for frailty, such as red blood cell count, IL-6, CRP, calcium, and hemoglobin (53). This method showed that the increase in the multivariate distance is accelerated with age, which represents the loss of integration of the system physiology. Similarly, the principal component analysis approach considered the variability of blood-based biomarkers, and consequently was showed to be an independent frailty predictor (54). Both methods included information from multiple systems to assess frailty, analogously to how CCM parameters were computed from HR and motor time-series, and how they were associated with physiological dysregulation and frailty status.

**Table 3.** Logistic models for predicting frailty using UEF motor score, HR dynamics, and CCM parameters. A significant association is represented by the asterisk.

| Independent variable | Parameter estimate | Standard error | Chi-square ($\chi^2$) | p-value (95% CI) |
|---|---|---|---|---|
| **Model 1 – UEF motor score** (AUC=0.78; AICc=53.94; Sensitivity=0.75; Specificity=0.75) | | | | |
| Intercept | 0.61 | 0.73 | 0.70 | 0.4 (-0.81:2.12) |
| UEF motor score | -0.05 | 0.02 | 6.85 | <0.01 (-0.08:-0.01)* |
| **Model 2 – HR dynamics** (AUC=0.84; AICc=44.25; Sensitivity=0.80; Specificity=0.75) | | | | |
| Intercept | -4.91 | 1.27 | 14.97 | <0.001(-7.92:-2.81)* |
| HR percent increase | 0.25 | 0.08 | 10.38 | <0.001 (0.12:0.44)* |
| **Model 3 – CCM** (AUC=0.74; AICc=55.60; Sensitivity=0.75; Specificity=0.25) | | | | |
| Intercept | 4.21 | 1.52 | 7.72 | <0.01 (1.77:7.86)* |
| CCM HR→M | -4.53 | 2.11 | 4.59 | 0.03 (-9.45:-0.98)* |
| **Model 4 – Combined UEF & HR dynamics** (AUC=0.87; AICc=76.67; Sensitivity=0.82; Specificity=0.83) | | | | |
| Intercept | -3.21 | 1.55 | 4.28 | 0.04 (-6.68:-0.45)* |
| HR percent increase | 0.23 | 0.08 | 7.73 | <0.01 (0.09:0.42)* |
| UEF motor score | -0.03 | 0.02 | 2.67 | 0.10 (-0.07:0.01) |
| **Model 5 – Combined UEF, HR dynamics & CCM** (AUC=0.91; AICc=42.80; Sensitivity=0.89; Specificity=0.83) | | | | |
| Intercept | 6.26 | 2.83 | 4.88 | 0.03 (1.69:13.14)* |
| HR percent increase | 0.22 | 0.08 | 7.38 | <0.01 (-0.42:-0.08)* |
| UEF motor score | -0.03 | 0.02 | 2.69 | 0.10 (-0.00:0.08) |
| CCM HR→M | 4.54 | 2.90 | 2.45 | 0.12 (-11.27:0.51) |

*HR heart rate; UEF upper-extremity function; CCM convergent cross-mapping; AUC area under curve; CI confidence interval; AICc Akaike's information criterion with correction for small sample size.*

*Frailty identification using multimodal models*

Current findings confirmed that assessing two physiological systems of motor and cardiac autonomic control, and especially the dynamic interaction between them, can improve frailty identification compared to models that focus on individual physiological systems in isolation. These two physiological systems were selected in this study as they are strongly associated with frailty. Muscle loss and weakness (sarcopenia and dynapenia) are the main symptoms of frailty, caused by inflammatory, metabolic, and hormonal derangements (2–10). Motor deficits and muscle weakness are commonly assessed using walking speed or grip strength tests (Fried phenotype) or counting deficits/disorders (Rockwood deficit index) (55,56). Nevertheless, performing walking tests in the clinical setting is cumbersome, and many frail older adults have walking disabilities. Grip strength, on the other hand, only measures muscle strength and cannot reveal other aspects of motor deficits. We have previously validated the sensor based UEF motor task to accurately detect systematic decrements in motor function associated with frailty, including slowness, weakness, inflexibility, fatigue, and motor variability (35,57).

In addition to the motor system, the implemented method included cardiac autonomic control. Previous research showed an association between frailty and an impaired ANS because of alterations in electrical conduction and action potential morphology (58,59). The presence of a compromised neurohormonal homeostasis associated with ANS dysfunction is, in turn, associated with health complications (60,61). HRV (i.e., variability in RR intervals within QRS-waves) during resting have been used for assessing ANS dysfunction, and has been proposed as a "vital sign" (62–64). However, between-subject and diurnal variability exists in resting HRV (e.g., due to breathing regulation and environmental factors (65–67)). Here, a novel measure of *HR dynamics* (HR response to physical activity) was introduced as a direct measure of sympathetic (during activity) and parasympathetic (during recovery) performance. The advantages of HR dynamics over HRV are twofold: 1) by normalizing the HR response to the resting condition, we will reduce between-subject and diurnal variability (17); and 2) we directly assess ANS performance and cardiac physiological reserve in response to a controlled stressor (physical task), to establish a stress-response model that can be further used for assessing interconnection measures.

As the last component, within the current study, we investigated the interconnection between physiological systems in response to stress caused by the physical task. The concept of stress-response testing for quantifying frailty has become the subject of recent research. Evidence suggests that differences in physiological reserve between non-frails and frails are subtle under the basal condition (68). Implementing provocative testing accentuates frailty-related alterations in measurable dynamics of physiological systems in response to stimuli. The provocative UEF test is designed to be hard enough to stress motor and cardiac systems, and not too demanding, so they can be incorporated in a routine clinical setting for frail older adults, especially those with walking disabilities. Simultaneous assessment of motor and heart function in this manner allows us to accurately quantify the dysregulation of interconnection between these systems. Further, the motion artifacts are minimum with the proposed testing, with HR measurement acquiring from the left side while the participant perform the physical task on the right side.

*Limitations and Further work*

Despite the promising findings of the current study, there are some limitations related to the recruited sample. First, the sample of community dwelling older adult selected for this study was small. Second, there was a limited number of frail participants, and therefore, pre-frail and frail groups were merged. Third, participants with arrhythmia and those who require β-blockers and pacemakers were excluded from the study. Also, test-retest reliability of CCM parameters were not investigated here. Therefore, the interconnection analysis should be confirmed in larger studies incorporating test-retest reliability measures. Additionally, we used time-series library lengths that may not provide accurate results for some participants, since some HR data may have a higher level of short-term complexity, leading to less dense attractor shadow manifolds and consequently a non-completely developed convergence of CCM parameters. Possible solutions would be to perform longer arm tests; however, this will lead to more physical demand on frail older adults.

*Conclusions and clinical implications*

In the present work a novel quantification of interconnection between motor and cardiac autonomic systems was implemented for frailty assessment. We demonstrated that CCM parameters showed weaker interconnection between motor and cardiac systems among pre-frail/frail older adults compared to non-frails. The new CCM parameters also showed promising results in improving frailty prediction within logistic models. The simplicity of the investigated UEF test permits performing it even for hospitalized bed-bound patients, for predicting therapy complications, in-hospital outcomes, and rehabilitation strategies. We expect to present this multimodal test as an alternative to accurate but impractical frailty assessment tools such the Fried phenotype, when patients are not able to walk. Further, commercialized wearable devices are now allow accurate assessment of HR and motion. Showing the proof of concept in the current study, in our future investigation, we will develop an easy-to-use app for Smart Watch for identifying frailty using simultaneous measures of motor and cardiac functions.

## Data availability

The datasets used for the current study are available from the corresponding author on reasonable request.

## Acknowledgements

This project was supported by two awards from the National Institute of Aging (NIA/NIH - Phase 2B Arizona Frailty and Falls Cohort 2R42AG032748–04 and NIA/NIH - 1R21AG059202-01A1). The views represented in this work are solely the responsibility of the authors and do not represent the views of NIH. We want to thank Ben Carpenter and Kayleigh Ruberto for their contribution to data collection and analysis.